\documentstyle[12pt]{article}

\begin{document}

\baselineskip=7mm

\begin{large}

\begin{center}
 {Generalization of the $N_pN_n$ Scheme and the Structure of the Valence Space}

\vspace{0.1in}
    
    Y. M.  Zhao$^{1}$, R. F. Casten$^{2}$,   and A. Arima$^3$

\vspace{0.2in}
$1$) Cyclotron Laboratory, the Institute of Physical and Chemical Research (RIKEN),
Hirosawa 2-1, Wako-shi,

 Saitama, 351-0198 Japan

$2$) A. W. Wright Nuclear Structure Laboratory,
Yale University, New Haven, CT 06520 USA

$3$)The House of Councilors, 2-1-1 Nagatacho,

   Chiyodaku, Tokyo 100-8962,  Japan

\vspace{0.315in}

\end{center}
\end{large}

\vspace{0.1in}

 The $N_pN_n$ scheme, which has  been extensively applied to 
even-even nuclei, is found to be  a very good benchmark for  
 odd-even, even-odd, and doubly-odd nuclei as well.  There are no 
apparent shifts in the correlations for these four classes of nuclei.
The compact correlations highlight the deviant behavior of the
Z=78  nuclei, are used
to deduce effective valence proton numbers near Z=64, 
and to study the evolution of the Z=64 subshell gap.

PACS Number(s):  27.60+j, 27.70+q, 27.80+w, 27.80+b
\vspace{0.4in}

\newpage

    Many physical systems, including atoms, nuclei and metallic clusters, 
exhibit shell structure.  Indeed, eigenvalues of the three-dimensional 
Schroedinger equation will tend to cluster in energy groupings 
(characterized by specific sets of principal and angular momentum quantum 
numbers)  for any reasonable central potential.  In the treatment of complex 
finite many-body systems, a common simplification is to invoke a "mean field" 
ansatz, replacing the sum of all the two-body interactions by a one-body 
potential.  Generally, such a procedure is only an approximation and various 
residual interactions need to be incorporated.  These will alter the 
predictions of the independent particle picture and may even lead to a 
breakdown of the shell structure, shell closures, and shell gaps. 

   Nuclei provide an ideal venue to study shell structure
and residual interactions   since they are
finite-body systems where the effective number of active bodies (the valence
nucleons) is generally quite small (0-30, say) and where one can both count and
change this number of bodies (the mass number)  in a controlled way.
Here, we wish to explore the evolution of collective behavior in nuclei
and the associated evolution of shell structure using an empirical correlation
scheme of collective observables that stresses the importance of the valence
residual p-n interaction.

The importance of the proton-neutron interaction in determining the
evolution of nuclear structure was   emphasized      long-ago by de Shalit and
Goldhaber \cite{Shalit}, and Talmi \cite{Talmi}. Two decades ago, Federman and
Pittel \cite{Pittel} emphasized that the driving mechanism in the
development of 
nuclear deformation is the proton-neutron interaction between nucleons in
  spin-orbit partner orbits. If the proton-neutron interaction
is a   controlling factor in the determination of nuclear structure,
a reasonable estimate of this interaction ought to be a useful
systematizing parameter with which the evolution of structure
could be correlated. 

In 1985 Casten described the $N_pN_n$ scheme  for
even-even nuclei \cite{Casten1}, in which 
 $E_{2_1^+}, E_{4_1^+}/E_{2_1^+}$, and $B(E2, 0_1^+\rightarrow 2_1^+)$ values  
were plotted against the product of valence proton number and
valence neutron number,  $N_pN_n$. 
The systematics for each observable  is 
very smooth, and similar from region to region.
It was found that the quantity $N_pN_n$ provides an excellent
scaling factor that allows one to assess the rapidity
of different transition regions and to predict the properties of
new nuclei \cite{Casten3}. Moreover, 
the slopes of different observables plotted against $N_pN_n$ are related 
to the average interaction, per proton-neutron pair, in the  highly
overlapping orbits whose occupation induces structural change.

However, most papers related to the $N_pN_n $ scheme have
 concentrated on the   
even-even case where there is  a rich array of compiled nuclear data.
It is therefore important  to see
whether the $N_pN_n$ scheme works, and how well it works,  in 
odd-A  and  doubly odd nuclei.
  The $N_pN_n$ concept  is more difficult to apply to 
    odd-A and odd-odd cases because 
 there can be a  very strong interplay 
between  collective and  single particle excitations,    
and  the low-lying excitation structures themselves are more complicated.
Moreover, adjacent nuclei differ in ground state and low-lying $J^{\pi}$
values so it is sometimes not clear which data to use in a systematic 
comparision. Finally, 
 observables   related to  odd-A nuclei and odd-odd nuclei are 
in general less well, and less  systematically, known than 
those of  even-even nuclei. 

The most extensive studies for odd-A nuclei to date have been
for the A=80-100 region.
In \cite{Bucu1} the 
$N_pN_n$ scheme was applied to both even-even
and odd-A nuclei in the A$\sim$80 
region; in  \cite{Bucu2} a few odd-A nuclei
with A$\sim$100 were considered;
in  \cite{Bucu3}, 
it was shown that  states
based on different single-particle excitations 
behave differently with $N_pN_n$. 
However, there has  not yet been any concerted effort 
towards a unified $N_pN_n$ treatment for   even-even,  odd-A  and
doubly odd nuclei over large mass regions.

It is therefore the purpose of this Letter to show  for the first time   
 that the simple $N_pN_n$ scheme 
 works equally  well for large regions of medium-heavy nuclei for 
 even-even, odd-A and the doubly-odd nuclei.
This extension to the $N_pN_n$ scheme  will significantly expand
its usefulness  for interpreting the sparse
data  soon-to-be-obtained on exotic nuclei far from stability.
We will also use these results to extract effective valence
proton numbers near Z=64 and N=83-91 in order to study the
breakdown of the Z=64 shell gap in even, odd and odd-odd nuclei. 

 We proceed by studying the deformation
parameter $e_2$ against $N_pN_n$. 
 The $e_2$ values 
 are taken from the macroscopic-microscopic calculations of \cite{Moller} for
 nuclei with known ground and excited states.
These deformations act  as   surrogates for directly  measured observables,
and therefore allow us to compare even and odd Z and N nuclei on the
same footing. 
These calcualtions are highly refined, and widely used.  For nuclei in or near 
the valley of stability, such as those considered here, they should provide an
excellent guide to realistic deformations, although it would be useful to
check them by experiment.   Of course, far from stability, the importance of
various residual interactions changes, as does the mean field itself, and
hence care should be taken in extending these results to new regions.  In any
case, for known nuclei, we believe
 that  the approximations used in \cite{Moller} are
reasonably good individually, and fully adequate for a systematic study
in large regions. Moreover, by using the deformation
rather than excitation energies to gauge the structure, one
avoids problems  with comparing levels with different spins.

 In Fig. 1, we present the quadrupole 
 deformation parameter in the Nilsson perturbed-spheroid parameterization,
  $e_2$, vs.  $N_pN_n$ for the nuclei in four different regions  
  ranging from Z=50 to 104, 
  namely the  50$<$Z$\le$66, 82$<$N$\le$104 region,
  the 66$<$Z$<$82,  82$<$N$\le$104  region,
 the 66$<$Z$<$82, 104$<$N$<$126   region, and the   82$<$Z$\le$104,
 126$<$N$<$155  
 region. The correlation between 
  $e_2$ and $N_pN_n$ is extraordinarily compact    
not only for the even-even nuclei but also for the 
  even-odd, odd-even and odd-odd cases as well (see solid symbols
  in Fig. 1a) and the full set of points in Figs. 1b, c, d). Moreover, 
   the correlations  are independent of the  
   even-even, even-odd, odd-even or odd-odd
  nature of the nuclei considered. No discernible bias for these
  classes of nuclei is visible  except for a slight
  difference between the points for even-proton number and
  odd-proton number for $N_pN_n$ values  less than 50 in Fig. 1c).

Among the  correlations shown,  Fig. 1a) 
 shows a greater broadening near $N_pN_n$  $\sim$ 50-100  
 than the other regions. This is a region   where  there is a
 subshell at Z=64  which we did not  take into account.
That is, we  used the proton magic numbers 50
and 82 for all  nuclei.
Below, we will examine the validity of these choices. To
facilitate that discussion,  Fig. 1a) uses open symbols 
for nuclei with N$\le91$ and $59\le$Z$\le$66. 
Another interesting point in  Fig. 1a) is that
there are  a number of data points with  $e_2$=0, which correspond to 
 the  N$=84$ isotones. These 
 isotones are very soft, which means that the shallow 
part of the potential energy against the deformation parameter  is wide. 
Hence there can be a large difference between 
the equilibrium deformation and the expectation    value of the deformation.  

In  b) and c) of Fig. 1,  several data points clearly
stand out to the upper left of  the  correlations.   Nearly all
have  Z=78 (Pt) and  lie in a complex region with large  
  $\gamma$-softness, 
 oblate shapes,  prolate shapes, and   transition regions     
between them. 
Nevertheless, other regions also show sharp shape changes but are  
not anomalous in the $N_pN_n$ plots. Therefore, it is worth
further effort to understand the behavior of the Z=78 Pt region and
whether these anomalous points reflect a different role for the p-n
interaction in these nuclei or a shortcoming in the calculated deformations
in  \cite{Moller}. 

While the concept of the valence space is important in understanding the 
structure of nuclei, in many cases the conventional counting of valence 
protons and neutrons is inadequate.  For example, near A=100 and 150, the 
Z=40 and 64 proton numbers take on magic character for certain neutron 
numbers but not for others \cite{Casten}.  Likewise, the neutron number N=20
is no longer magic for the
neutron rich nucleus $^{32}$Mg \cite{Motobayashi}.
Indeed, it is expected that magicity may
well be a fragile construct far from stability.
This fragility is a result both of changes to the mean field and
 to the valence p-n residual interaction\cite{england,Heyde}.
Its effects might be expected to show up in the $N_pN_n$ scheme.  Indeed, in 
even-even nuclei, effective $N_p$ values have been discussed for both the A=100 
and 150 regions [4,14-17].

The present results  give us the opportunity to probe this issue more deeply, 
by extracting effective $N_p$ values in the A=150 region from even, odd and 
odd-odd nuclei simultaneously and in a unified way.

In Fig. 1a), the solid symbols are for the  59$\le$Z$\le$66 
and N$\ge$92 nuclei, and all nuclei with 50$<$Z$\le$58.  They form an
extraordinarily compact trajectory, while the 59$\le$Z$\le$66  and N$\le$91  
nuclei deviate
strongly to the right.  This arises because, for these latter nuclei, Z=64 
acts as a magic or partially magic number whereas Fig. 1a) was constructed 
using Z=50 as magic.  Hence these nuclei were plotted at inappropriately large 
$N_pN_n$ values.  The opposite assumption, that Z=64 is magic for N$\le$91 
is also too extreme.  As shown in Fig. 2, this leads to 
an overshoot of these points to the left.

Clearly, by assuming the validity of the compact correlation for nuclei not
affected by a Z=64 gap, that is those marked by solid symbols in Fig. 1a),
and shifting the ``deviant" nuclei leftward to this correlation, we can extract the
effective $N_p$ values for these nuclei and thereby assess the breakdown and
dissolution of the Z=64 gap. Equivalently, one can
shift the anomalous data points in Fig.2 to the right. 
 The process is similar to that used in \cite{Casten1}
for even-even nuclei but now is extended uniformly to all species.

Fig. 3 illustrates how this approach works by looking at a subset of
the points in Fig.2$--$those for even-odd nuclei.
Here, the solid symbols are
the nuclei unaffected by a Z=64 gap. The open symbols lie at various distances
from the main correlation:  consistently,  the Z=64, 66 isotopes lie farthest, and
the Z=62, and 60 isotopes occur successively closer.
The amount of shifting required for each point is determined by fitting 
an exponential function to the normal (solid symbol) data in Fig.3,  and
such a fitting curve is used as a guide  to deduce 
the appropriate $N_p$ value for that $e_2$.
The  resulting effective 
$N_p$ values for all the data of 
 Fig. 1a)  are summarized in Table 1 and shown in Fig. 4. They are given
 in the Table to the nearest odd(even) integers for  odd(even)-Z nuclei.
Note that in Table 1 we do not present effective valence proton numbers for
the N=84 isotones  since, as discussed above, these nuclei are soft and the equilibrium and
mean deformations may differ considerably,  and also the calculated
deformations can be very sensitive to small perturbations.  The results in
Table 1 
demonstrate a gradual breakdown of the Z=64 shell gap, which accelerates near
N=90, and consistency regardless of whether the nuclei are even-even,
odd-even, even-odd, or odd-odd.

To summarize, the $N_pN_n$ scheme,  which has  been extensively
studied for  even-even nuclei,  is found to be equally
applicable to all species of  medium-heavy 
nuclei: even-even, odd-even, even-odd, and odd-odd.
The $N_pN_n$ correlations are not sensitive to the odd-even difference. 
This supports the idea  that the proton-neutron
interaction plays a similar role 
regardless of the even-odd character of the nuclei, and 
 suggests that the average strength of 
the valence proton-neutron interaction is almost constant between  
even-even and their odd-A/odd-odd neighbors. The extremely compact $N_pN_n$
trajectories highlight a few deviant nuclei. Finally, effective valence
proton numbers were extracted from these correlations and found to be
 also insensitive to 
the category of nucleus. This gives a deeper view of the breakdown of
the Z=64 magicity near neutron number 90.

The present work 
extends the realm of application of the $N_pN_n$ scheme 
 to all types of  nuclei.  Given that compact correlation schemes,
such as $N_pN_n$, magnify anomalous behavior (e.g., 
the Z=78 nuclei discussed above), and probe the valence space (i.e.,
the effective valence nucleon numbers), the present results and approach can
  provide a
more general tool to disclose new and different types of shell structure or 
structural evolution (e.g., changes in shell structure and magicity)
in exotic nuclei. 

{\bf ACKNOWLEDGEMENTS}
     
Discussions with  Drs. N.V. Zamfir,  P. Moeller,
 N. Yoshinaga,   S. Yamaji, and  S. G. Zhou are
appreciated  gratefully. 
 One of the authors (YMZ) acknowledges the
Science and Technology Agency of Japan (Contract: 297040) for supporting
this project. Work supported in part by the U.S. DOE under grant number
DE-FG02-91ER40609.

\newpage

Captions:

\vspace{0.2in}

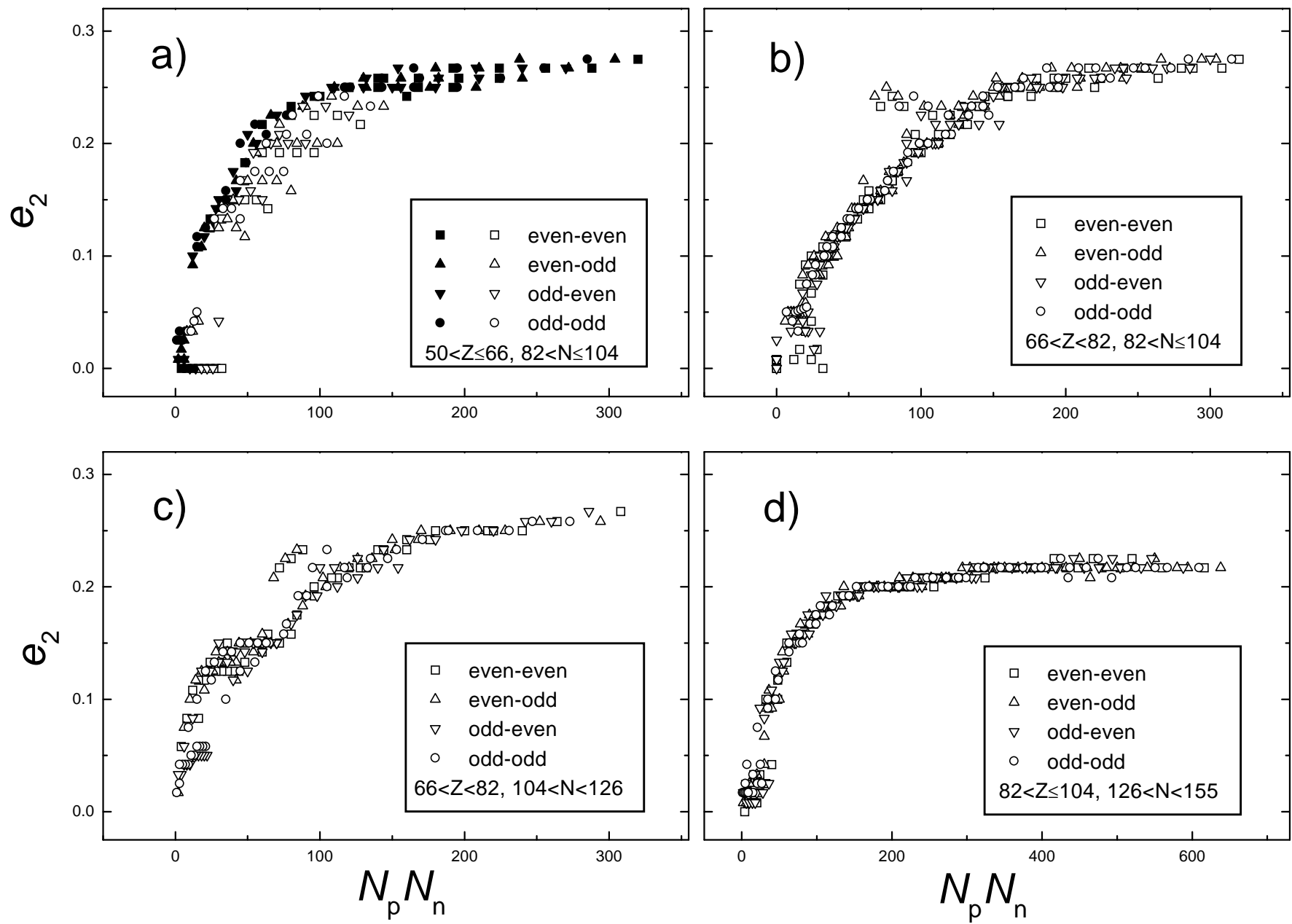
{FIG. 1. The deformation parameter $e_2$ vs. $N_pN_n$. a)~~ for nuclei 
with   50$<$Z$\le$66 and 82$<$N$\le$104. Open symbols for Z=59-66 and
N$\le$91. Solid symbols for all other nuclei (i.e., 50$<$Z$\le$58 for
all neutron numbers and 59$\le$Z$\le$66 for N$\ge$92.); b) ~~
66$<$Z$<$82 and  82$<$N$\le$104; c) ~~
 66$<$Z$<$82 and  104$<$N$<$126; d) ~~
 82$<$Z$\le$104  and  126$<$N$<$155.
Note the scale
change in part d) to accomodate the larger $N_pN_n$ values in this mass region. }

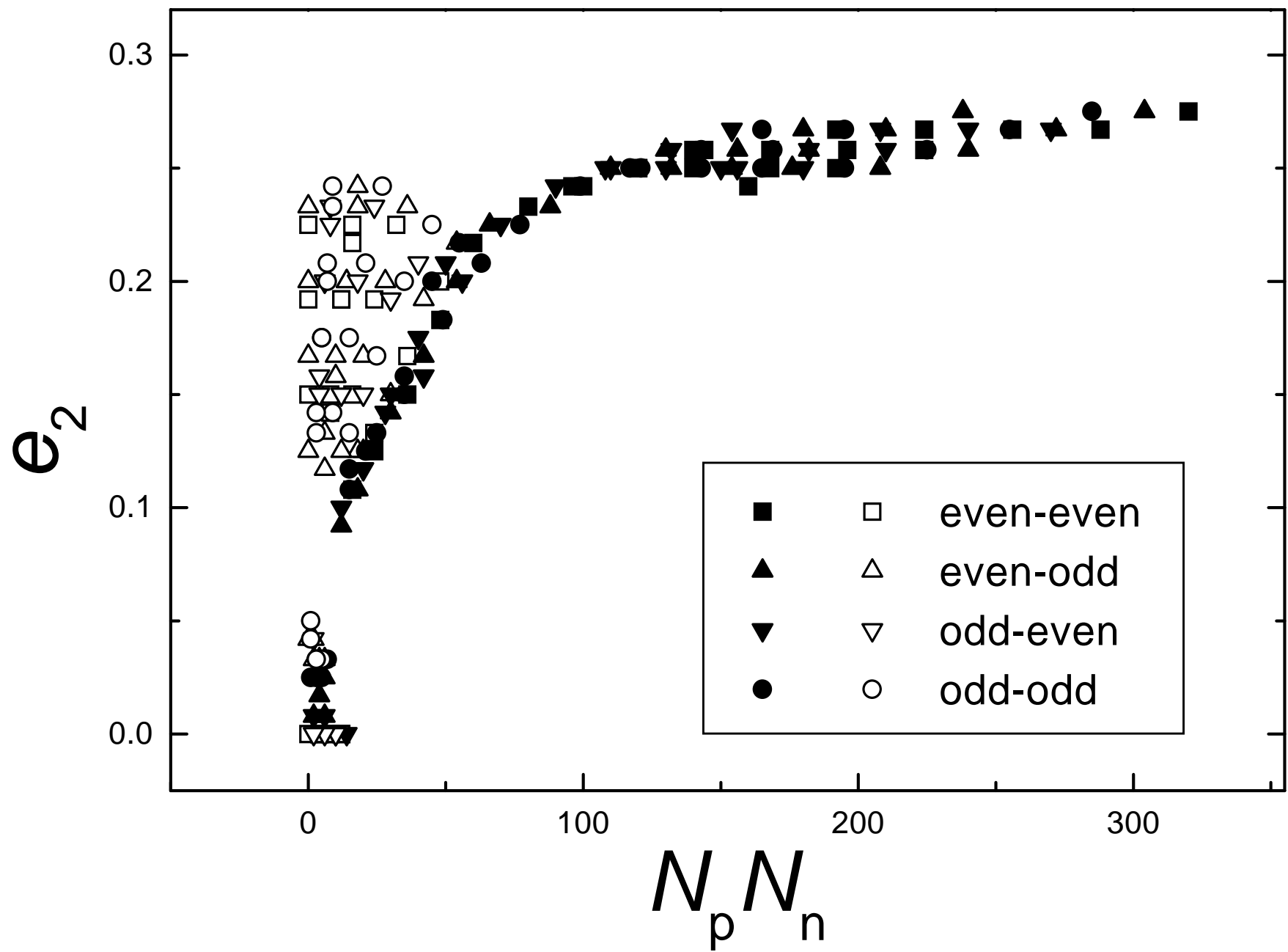
 {FIG. 2. Similar to Fig. 1a) except Z=64 is used as a magic number instead
 of 82 for N$\le$91. }

\vspace{0.2in}

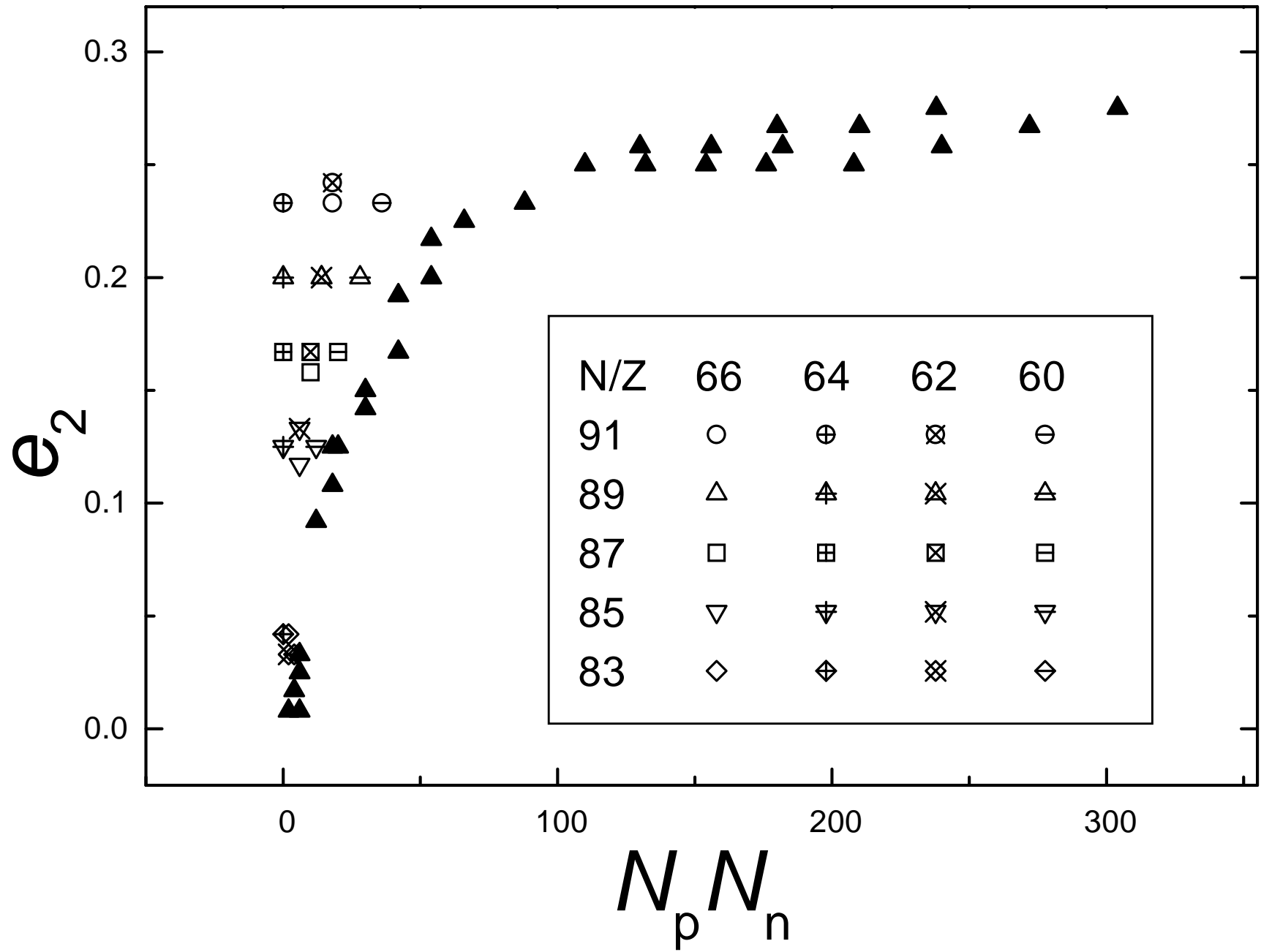
{FIG. 3. Extract from Fig. 2 for even-odd nuclei, where different symbols 
are used to denote nuclei with 60$\le$Z$\le$66 and  N$\le$91.}

\vspace{0.2in}

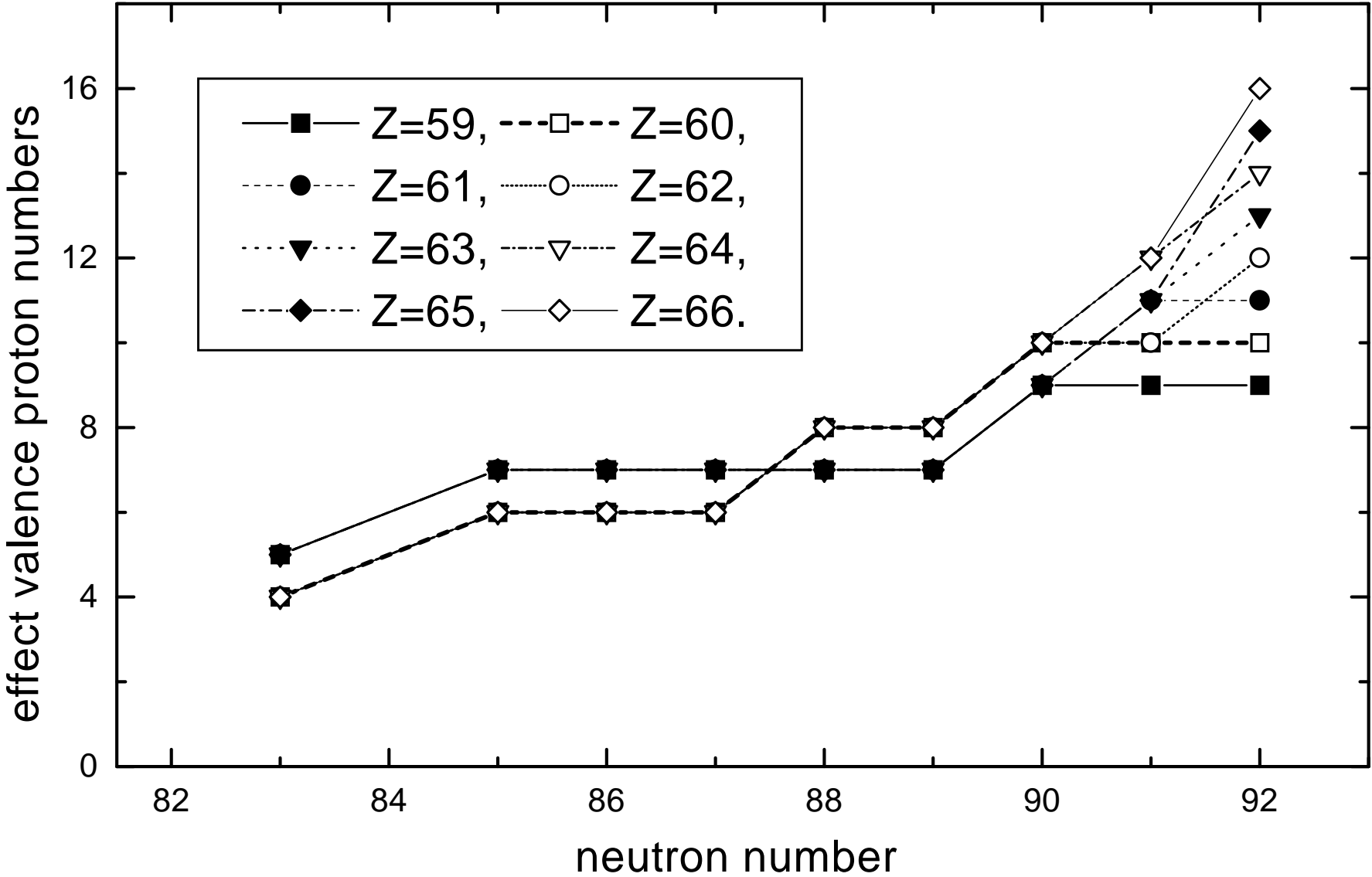
{FIG. 4. Summary of the  effective valence proton numbers obtained in this
work. }

\vspace{0.5in}

\begin{table}
\caption{Effective proton numbers for nuclei near the Z=64 subshell.}
\begin{tabular}{cccccccccc} \hline \hline
Z$/$N & 83  & 85 & 86  & 87 & 88 & 89 & 90& 91 & 92\\  \hline
59    & 5   & 7  & 7   & 7  & 7  & 7  & 9 & 9  & 9  \\
60    & 4   & 6  & 6   & 6  & 8  & 8  & 10& 10 & 10 \\
61    & 5   & 7  & 7   & 7  & 7  & 7  &  9& 11 & 11 \\
62    & 4   & 6  & 6   & 6  & 8  & 8  & 10& 10 & 12 \\
63    & 7   & 7  & 7   & 7  & 7  & 7  & 9 & 11 & 13 \\
64    & 4   & 6  & 6   & 6  & 8  & 8  & 10&  12& 14 \\
65    & 5   & 7  & 7   & 7  & 7  & 7  & 9 & 11 & 15 \\
66    & 4   & 6  & 6   & 6  & 8  & 8  & 10&  12& 16 \\   \hline \hline
\end{tabular}
\label{two}
\end{table}


\newpage


\begin{thebibliography}{50}
\bibitem{Shalit} A. De Shalit and M. Goldhaber, Phys. Rev. {\bf 92}, 1211(1953). 

\bibitem{Talmi} I. Talmi, Rev. Mod. Phys. {\bf 34}, 704(1962).

\bibitem{Pittel}  P. Federman and S. Pittel, Phys. Lett. {\bf 69B}, 385(1977);
{\bf 77B}, 29(1977); Phys. Rev. C{\bf 20}, 820(1979);
  P. Federman, S. Pittel, and R. Campas, Phys. Lett. {\bf 82B}, 9(1979). 


\bibitem{Casten1}  R. F. Casten, Phys. Rev. Lett. {\bf 54}, 1991(1985); 
 Nucl. Phys. {\bf A443}, 1(1985);  Phys. Rev. Lett. {\bf 58},  658(1987). 

\bibitem{Casten3} For a recent review, see R. F. Casten and N. V. Zamfir,  J. Phys. G {\bf 22}, 1521(1996).


\bibitem{Bucu1} S.L. Tabor, Phys. Rev. C{\bf 34}, 311(1986).

\bibitem{Bucu2} H.Dejbakhsh,  Phys. Lett. B {\bf 210}, 50(1988). 

\bibitem{Bucu3} D. Bucurescu et al., 
 Phys. Lett. B {\bf 229}, 321(1989). 

\bibitem{Moller}  P. Moeller, J. R. Nix, W. D. Myers, and
W. J. Swiatecki,      Atomic Data Nucl. Data Tables {\bf 59}, 185(1995).


\bibitem{Casten}  R. F. Casten et al.,  Phys. Rev. Lett. 47, 1433(1981). 

\bibitem{Motobayashi} T.Motobayashi et al., 
Phys. Lett. B{\bf 346}, 9(1995).


\bibitem{england} J. Dobaczewski and W. Nazarewicz,
Phil. Trans. R. Proc. Lond. {\bf A356}, 2007(1998).


\bibitem{Heyde} K. Heyde et al., 
Phys. Lett. B{\bf 155}, 303(1985).


\bibitem{Z=64} Y. M. Zhao and Y. Chen, Phys. Rev. C{\bf 52}, 1453(1995);

\bibitem{Wolf} A. Wolf and R. F. Casten, Phys. Rev. C{\bf 36}, 851(1987). 

\bibitem{Scholten} O. Scholten, Phys. Lett. {\bf 127B}, 144(1983).


\bibitem{Chuu} S. T. Hsieh et al., J. Phys. G{\bf 12}, L167(1986);
D. S. Chuu et al., Nucl. Phys. {\bf A482}, 679(1988);
C. S. Han et al., Phys. Rev. C{\bf 42}, 280(1990). 



\end{thebibliography}
\end{document}